\documentstyle[12pt]{article}

\def\hybrid{\topmargin -25pt    \oddsidemargin 0pt
        \headheight 0pt \headsep 0pt
        \textwidth 6.25in       
        \textheight 9.5in       
        \marginparwidth .875in
        \parskip 5pt plus 1pt   \jot = 1.5ex}
\def\cQ{{\cal Q}}
\def\cG{{\cal G}}
\def\cL{{\cal L}}
\def\cH{{\cal H}}
\def\ket#1{|{#1}\rangle}
\def\noi{\noindent}
\def\half{{1\over2}}
\def\baselinestretch{1.2}

\catcode`\@=11

\def\marginnote#1{}
\def\draftlabel#1{{\@bsphack\if@filesw {\let\thepage\relax
   \xdef\@gtempa{\write\@auxout{\string
      \newlabel{#1}{{\@currentlabel}{\thepage}}}}}\@gtempa
   \if@nobreak \ifvmode\nobreak\fi\fi\fi\@esphack}
        \gdef\@eqnlabel{#1}}
\def\@eqnlabel{}
\def\@vacuum{}
\def\draftmarginnote#1{\marginpar{\raggedright\scriptsize\tt#1}}

\def\draft{\oddsidemargin -.2truein
        \def\@oddfoot{\sl preliminary draft \hfil
        \rm\thepage\hfil\sl\today\quad\militarytime}
        \let\@evenfoot\@oddfoot \overfullrule 3pt
        \let\label=\draftlabel
        \let\marginnote=\draftmarginnote
   \def\@eqnnum{(\theequation)\rlap{\kern\marginparsep\tt\@eqnlabel}%
\global\let\@eqnlabel\@vacuum}  }

\def\preprint{\twocolumn\sloppy\flushbottom\parindent 2em
        \leftmargini 2em\leftmarginv .5em\leftmarginvi .5em
        \oddsidemargin -.5in    \evensidemargin -.5in
        \columnsep .4in \footheight 0pt
        \textwidth 10.in        \topmargin  -.6in
        \headheight 12pt \topskip .3in
        \textheight 6.9in \footskip 0pt
        \def\@oddhead{\thepage\hfil\addtocounter{page}{1}\thepage}
        \let\@evenhead\@oddhead \def\@oddfoot{} \def\@evenfoot{} }

\def\numberbysection{\@addtoreset{equation}{section}
        \def\theequation{\thesection.\arabic{equation}}}

\def\underline#1{\relax\ifmmode\@@underline#1\else
        $\@@underline{\hbox{#1}}$\relax\fi}

\def\titlepage{\@restonecolfalse\if@twocolumn\@restonecoltrue
\onecolumn
     \else \newpage \fi \thispagestyle{empty}\c@page\z@
        \def\thefootnote{\fnsymbol{footnote}} }

\def\endtitlepage{\if@restonecol\twocolumn \else \newpage \fi
        \def\thefootnote{\arabic{footnote}}
        \setcounter{footnote}{0}}  

\catcode`@=12
\relax

\def\figcap{\section*{Figure Captions\markboth
        {FIGURECAPTIONS}{FIGURECAPTIONS}}\list
        {Figure \arabic{enumi}:\hfill}{\settowidth\labelwidth{Figure
999:}
        \leftmargin\labelwidth
        \advance\leftmargin\labelsep\usecounter{enumi}}}
 \relax
\def\tablecap{\section*{Table Captions\markboth
        {TABLECAPTIONS}{TABLECAPTIONS}}\list
        {Table \arabic{enumi}:\hfill}{\settowidth\labelwidth{Table
999:}
        \leftmargin\labelwidth
        \advance\leftmargin\labelsep\usecounter{enumi}}}
 \relax
\def\reflist{\section*{References\markboth
        {REFLIST}{REFLIST}}\list
        {[\arabic{enumi}]\hfill}{\settowidth\labelwidth{[999]}
        \leftmargin\labelwidth
        \advance\leftmargin\labelsep\usecounter{enumi}}}
 \relax
\makeatletter
\newcounter{pubctr}
\def\publist{\@ifnextchar[{\@publist}{\@@publist}}
\def\@publist[#1]{\list
        {[\arabic{pubctr}]\hfill}{\settowidth\labelwidth{[999]}
        \leftmargin\labelwidth
        \advance\leftmargin\labelsep
        \@nmbrlisttrue\def\@listctr{pubctr}
        \setcounter{pubctr}{#1}\addtocounter{pubctr}{-1}}}
\def\@@publist{\list
        {[\arabic{pubctr}]\hfill}{\settowidth\labelwidth{[999]}
        \leftmargin\labelwidth
        \advance\leftmargin\labelsep
        \@nmbrlisttrue\def\@listctr{pubctr}}}
 \relax
\makeatother
\newskip\humongous \humongous=0pt plus 1000pt minus 1000pt

\newif\ifdtup

\font\Scbig=cmss10 scaled\magstep1
\font\Scscr=cmss8 scaled\magstep1
\font\Scscrscr=cmss8
\newfam\Scfam
\textfont\Scfam=\Scbig
\scriptfont\Scfam=\Scscr
\scriptscriptfont\Scfam=\Scscrscr
\def\Sc{\fam\Scfam}

\relax
\hybrid
\def\lvm{\leavevmode\hbox to\parindent{\hfill}}

\def\thefootnote{\fnsymbol{footnote}}
\def\BE{\begin{equation}}
\def\EE{\end{equation}}
\def\BA{\begin{eqnarray}}
\def\EA{\end{eqnarray}}

\def\a{\alpha}
\def\th{\theta}

\def\P{\Phi}

\def\e{\epsilon}

\def\tt{\bar\tau}

\def\lvm{\leavevmode\hbox to\parindent{\hfill}}
\def\bar{\overline}
\def\req#1{(\ref{#1})}

\def\L{\left}
\def\R{\right}

\def\BE{\begin{equation}}
\def\EE{\end{equation} \vskip 0.30\baselineskip}
\def\BA{\begin{array}}
\def\EA{\end{array}}

\def\noi{\noindent}

\def\frac#1#2{{\textstyle{{#1}\over{#2}}}}
\def\half{{1\over2}}

\def\Kr#1{\delta_{{#1},0}}
\def\ket#1{|{#1}\rangle}

\def\ccases#1#2{\L\{\!\new\BA{l}{#1}\\ {#2}\EA\R.}

\def\cA{{\cal A}}
\def\cG{{\cal G}}
\def\cH{{\cal H}}

\def\cL{{\cal L}}

\def\cQ{{\cal Q}}

\def\cU{{\cal U}}

\def\open#1{\mbox{{\bf{#1}}}}

\def\oZ{{\open Z}}

\def\ctop{{\Sc c}}
\def\htop{{\Sc h}}

\def\a{\alpha}
\def\b{\beta}
\def\g{\gamma}
\def\Ups{\Upsilon}

\def\ie{{\it i.e.}}

\def\Qz{\cQ_0}
\def\Gz{\cG_0}
\def\Qn{$\Qz$}
\def\Gn{$\Gz$}
\def\kc{{\ket{\chi}}}

\newif\ifold \oldtrue \def\new{\oldfalse}
\let\ssection=\section
\def\section{\setcounter{equation}{0}\ssection}

\begin{document}
\renewcommand{\theequation}{\thesection.\arabic{equation}}
\newcommand{\beq}{\begin{equation}}
\newcommand{\eeq}[1]{\label{#1}\end{equation}}
\newcommand{\ber}{\begin{eqnarray}}
\newcommand{\eer}[1]{\label{#1}\end{eqnarray}}
\begin{titlepage}
\begin{center}

\hfill IMAFF-96/38\\
\hfill NIKHEF-96-007\\
\hfill hep-th/9602166
\vskip .4in

{\large \bf  Interpretation of the Determinant Formulae for the 
Chiral Representations of the N=2 Superconformal Algebra}
\vskip .4in

{\bf Beatriz Gato-Rivera}$^{a,b}$ {\bf and Jose Ignacio Rosado}$^a$\\
\vskip .3in

${\ }^a$ 
{\em Instituto de Matem\'aticas y
 F\'\i sica Fundamental, CSIC,\\ Serrano 123,
Madrid 28006, Spain} \footnote{e-mail addresses:
bgato@pinar1.csic.es, jirs@pinar1.csic.es}\\
\vskip .25in

${\ }^b$
 {\em NIKHEF-H Kruislaan 409, NL-1098 SJ Amsterdam, The Netherlands}\\

\end{center}

\vskip .4in

\begin{center} {\bf ABSTRACT } \end{center}
\begin{quotation}

 We show that the N=2 determinant formulae of the
Aperiodic NS algebra and the Periodic R algebra can be
applied directly to incomplete 
Verma modules built on chiral primary states
and on Ramond ground states, respectively, provided
one modifies the interpretation of the zeroes in an appropriate
 way. That is, the zeroes of the determinat formulae account for
the highest weight singular states built on chiral primaries and
on Ramond ground states, but the identification of the levels and
relative U(1) charges of the singular states is different than
for complete Verma modules. In particular, half
of the zeroes of the quadratic vanishing surfaces $f^A_{r,s}=0$
and $f^P_{r,s}=0$ correspond to uncharged singular states,
and the other half correspond to charged singular states. 
We derive the spectrum of singular states
built on chiral primaries, including the singular states 
of the Twisted Topological algebra, and the spectrum of 
singular states built on the Ramond ground states.
We also uncover the existence of 
 non-highest weight singular states which are not secondary
 of any highest weight singular state.    

\end{quotation}
\vskip .15in

April 1996\\  
 
\end{titlepage}
\vfill
\eject
\def\baselinestretch{1.2}
\baselineskip 17 pt
\section{Introduction}\lvm

Various aspects concerning singular states\footnote{By singular state
we mean null or zero-norm state. However, we will use the term
singular state to denote highest weight singular states, unless
otherwise indicated.}
 of the Twisted Topological
N=2 Superconformal algebra have been studied in several papers during
the last four years \cite{BeSe2}, \cite{BeSe3}, \cite{Sem},
 \cite{BJI2}, \cite{BJI3}, \cite{SemTip}. In most of those papers
some explicit examples of topological
 singular states were written down,
ranging from level 2 until level 4. In addition, in \cite{Sem}
and \cite{SemTip} general formulae were given 
 for the spectra of the uncharged singular
states, that is, the spectra of U(1) charges
corresponding to the topological Verma modules which
contain uncharged singular states. Although the formulae fitted
 with the few known data, a proof or derivation was
 lacking. 

One of the aims of this paper is precisely to derive the spectra
of U(1) charges for all the four kinds of topological singular
states \cite{BJI6};
 that is, the uncharged \Qn -closed (BRST-invariant)
$\kc^{(0)Q}$, the uncharged \Gn -closed (anti-BRST invariant)
$\kc^{(0)G}$, the  charge $(-1)$ \Qn -closed $\kc^{(-1)Q}$ and
the charge $(+1)$ \Gn -closed $\kc^{(1)G}$. 
 The idea is to derive these spectra from the spectra of singular
states of the untwisted Aperiodic NS algebra, \ie\
 using the corresponding determinant formula\footnote{The N=2
 determinant formulae have been
checked by several authors (see for example \cite{Nam},
\cite{Kir1}, \cite{Kir2}, \cite{YuZh}
and \cite{KaMa3}) and therefore it seems unlikely that they
can contain any mistakes.}
written down for the first time by
Boucher, Friedan and Kent \cite{BFK} 
 However, 
the determinant formula for the Aperiodic NS algebra
does not apply  directly to incomplete Verma modules built
 on chiral\footnote{By chiral we mean chiral and antichiral,
unless otherwise indicated.}
 primaries, which are precisely the only meaningful
Verma modules of the Twisted Topological algebra.

A direct approach to the problem would be to compute
 specific ``chiral" determinant formulae for such
 representations. The approach that we follow here is simpler,
 although equivalent since we actually find
the zeroes of the chiral determinant formulae 
(they are contained in the zeroes of the determinant formula in
a simple, easy to identify pattern), and we also identify
 the levels and relative U(1) charges
of the singular states associated to them. 

 Although we do not
present rigorous proofs, our results agree with all the
known data: computation of NS singular states built on
chiral primaries from level 1/2 to level 3 \cite{BJI6} 
and also the data for levels 7/2 and 4, as deduced from the
 computation of the topological singular states at level 4
\cite{Sem}. In addition, the simplicity of our derivation 
strongly suggests that these results must hold at any level.

 Our starting point is the ansatz that the zeroes of the
determinant formula contain the zeroes of the
chiral determinant formulae (one determinat formula for
 chiral representations and another
for antichiral representations).
 The simplest assumption
is in fact that the zeroes of the chiral determinant formulae
coincide with the zeroes of the determinant formula specialized
to the cases $\Delta=\pm{\htop\over2}$, where $\Delta$ is the
conformal weight and $\htop$ the U(1) charge of the primary
state on which the Verma module is built.

 Then we impose 
the symmetries between the charged and
uncharged singular states dictated by the 
spectral flow automorphism of the Twisted Topological algebra
 \cite{BJI3}. To be precise, this automorphism, which maps 
topological singular states into each other,
 implies after the untwisting
 that the charged and uncharged
singular states of the Aperiodic NS algebra, built on
chiral primaries, must come in pairs with a precise
relation between the Verma modules to which they belong.
 This fact is accounted for by the determinant formula
 (under the simplest assumption)
if half of the zeroes of the quadratic vanishing surface
$f^A_{r,s}=0$ correspond to uncharged singular states, at level $rs\over2$,
and the other half correspond to charged singular states,
at level $r(s+2)-1\over2$
(in addition to the zeroes of the vanishing plane $g_k^A = 0$).
This contrasts sharply with the spectra of singular
states for complete Verma modules built
 on non-chiral primaries \cite{BFK},
for which  all the zeroes of the quadratic vanishing surface
correspond to uncharged singular states.

 The actual computation of NS singular states built on chiral 
 primaries agrees with this analysis,
 showing that indeed the zeroes of the chiral determinant formulae
coincide with the zeroes of the determinant formula specialized
to the cases $\Delta=\pm{\htop\over2}$. The interpretation of
the zeroes in terms of spectra of singular states, \ie\ levels
and relative charges, is different though. In particular,
 the spectrum for both charged and 
uncharged NS singular states built on chiral primaries
 is given by two-parameter expressions whereas for complete
Verma modules the spectrum of charged singular
 states is given by a one-parameter expression \ie\ the solutions to the
vanishing plane $g_k^A=0$. 

{}From these two-parameter expressions we get straightforwardly
the spectra of U(1) charges for the singular states of the
Twisted Topological algebra, as is explained in section 3.
Our expressions coincide with the expressions given in 
 \cite{Sem} and \cite{SemTip} accounting for the known data
(until level 4).  
    
We repeat the analysis for the Periodic R algebra for Verma modules
 built on the Ramond ground states. As expected, we
find the same results, \ie\  half of
the zeroes of $f^P_{r,s}=0$ are related to uncharged singular
states and the other half are related to charged singular
states. The reason is that
 the spectral flows map singular states of the Aperiodic
NS algebra built on chiral primaries into singular states of the
Periodic R algebra built on the Ramond ground states.

 We show therefore that the zeroes of the determinant formulae
 of the Aperiodic NS algebra and the Periodic R algebra account for
the singular states built on chiral primary states
and on Ramond ground states, respectively, but the identification
of the levels and relative charges of the singular states is
different than for complete Verma modules.

This paper is organized as follows.
In section 2 we discuss some basic facts about the singular
states of the Twisted Topological algebra and the relations 
between them. 
In section 3 we explain the direct relation between these and
the singular states of the Aperiodic NS algebra.
 We show that the untwisting of the \Gn -closed
topological singular states produces singular states of the
Aperiodic NS algebra in a one-to-one
way. As a consequence the charged and uncharged NS singular states built
on chiral primaries must come in pairs, although in different
Verma modules related to each other by the spectral flow
automorphism of the Twisted Topological Algebra.
We use this result in
 section 4, where we deduce the spectrum of U(1) charges $\htop$
 (and conformal weights $\Delta$) for the NS singular states
built on chiral primaries, just by analyzing the zeroes
of the determinant formula and comparing the different
possibilities with the actual
data of singular states. The derivation of the spectrum
of U(1) charges for the singular states of the Twisted Topological
algebra follows straightforwardly. 
In section 5 we repeat the analysis for the Periodic R algebra,
via the spectral flows, and we write down the spectrum of
U(1) charges for the
singular states built on the Ramond
ground states.
Section 6 is devoted to conclusions and final remarks.

Finally, in the Appendix we analyze thoroughly 
 the NS singular states at levels 1 and $3\over2$. 
 We write down all the equations 
resulting from the highest weight conditions, showing
that these equations, and therefore their solutions, 
are different when imposing or not chirality on the
primary state on which the singular states are built.
Namely, when the primary state is non-chiral all the 
solutions of the quadratic vanishing surface $f_{1,2}^A=0$
correspond to level 1 uncharged singular states. When the
primary state is chiral, however, half of the solutions
specialized to the cases $\Delta = \pm {\htop\over2}$
 correspond to level $3\over2$ charged singular states. 
We also show that these charged singular states become
non-highest weight singular states of a very special kind,
once the chirality on the primary state is switched off,
because then they are not secondary of any h.w. singular state.

\section{Singular States of the Twisted Topological Algebra}\lvm

The Twisted Topological algebra 
obtained by twisting the N=2 Superconformal
algebra  \cite{[EY]}, \cite{[W-top]}, \cite{DVV} reads

\BE\new\BA{lclclcl}
\L[\cL_m,\cL_n\R]&=&(m-n)\cL_{m+n}\,,&\qquad&[\cH_m,\cH_n]&=
&{\ctop\over3}m\Kr{m+n}\,,\\
\L[\cL_m,\cG_n\R]&=&(m-n)\cG_{m+n}\,,&\qquad&[\cH_m,\cG_n]&=&\cG_{m+n}\,,
\\
\L[\cL_m,\cQ_n\R]&=&-n\cQ_{m+n}\,,&\qquad&[\cH_m,\cQ_n]&=&-\cQ_{m+n}\,,\\
\L[\cL_m,\cH_n\R]&=&\multicolumn{5}{l}{-n\cH_{m+n}+{\ctop\over6}(m^2+m)
\Kr{m+n}\,,}\\
\L\{\cG_m,\cQ_n\R\}&=&\multicolumn{5}{l}{2\cL_{m+n}-2n\cH_{m+n}+
{\ctop\over3}(m^2+m)\Kr{m+n}\,,}\EA\qquad m,~n\in\oZ\,.\label{topalgebra}
\EE

\noi
where $\cL_m$ and $\cH_m$ are the bosonic generators corresponding
to the energy momentum tensor (Virasoro generators)
 and the topological $U(1)$ current respectively, while
$\cQ_m$ and $\cG_m$ are the fermionic generators corresponding
to the BRST current and the spin-2 fermionic current
 respectively. The ``topological central
charge" $\ctop$ is the true central charge of the N=2
Superconformal algebra \cite{Ade}, \cite{PDiV},
\cite{Kir1} \cite{LVW} .

This algebra is satisfied by the two sets of topological generators

\BE\new\BA{rclcrcl}
\cL^{(1)}_m&=&\multicolumn{5}{l}{L_m+\half(m+1)H_m\,,}\\
\cH^{(1)}_m&=&H_m\,,&{}&{}&{}&{}\\
\cG^{(1)}_m&=&G_{m+\half}^+\,,&\qquad &\cQ_m^{(1)}&=&G^-_{m-\half}
\,,\label{twa}\EA\EE

\noi
and

\BE\new\BA{rclcrcl}
\cL^{(2)}_m&=&\multicolumn{5}{l}{L_m-\half(m+1)H_m\,,}\\
\cH^{(2)}_m&=&-H_m\,,&{}&{}&{}&{}\\
\cG^{(2)}_m&=&G_{m+\half}^-\,,&\qquad &
\cQ_m^{(2)}&=&G^+_{m-\half}\,,\label{twb}\EA\EE

\noi
corresponding to the two possible twistings of the superconformal generators
$L_m, H_m, G^{+}_m$ and $G^{-}_m$. We see that $G^{+}$ and $G^{-}$ play
mirrored roles with respect to the definitions of $\cG$ and $\cQ$. In
particular $(G^{+}_{1/2}, G^{-}_{-1/2})$ results in
$(\cG^{(1)}_0, \cQ^{(1)}_0)$, while
 $(G^{-}_{1/2}, G^{+}_{-1/2})$ gives
$(\cG^{(2)}_0, \cQ^{(2)}_0)$, so that
 the topological chiral primary states
$\ket\P^{(1)}$ and $\ket\P^{(2)}$
(annihilated by both \Qn\ and \Gn)
 correspond to the antichiral and the chiral primary states of the
Aperiodic NS algebra respectively. We remind the reader that 
the chiral and the antichiral primary states of the Aperiodic NS
algebra are those primaries annihilated by $G^+_{-1/2}$
and $G^-_{-1/2}$ respectively.     

As usual, only the topological primary states
 that are chiral will be considered. The 
anticommutator $\{\cG_0,\cQ_0\}=2\cL_0$ shows that these 
states have zero conformal weight, therefore their only quantum number is 
their U(1) charge $\htop$ ($\cH_0\ket\Phi=\htop\ket\Phi$).
This anticommutator also shows 
that a $\cQ_0$-closed secondary state is $\cQ_0$-exact as well, and 
similarly with $\cG_0$. Therefore, the only possible physical states 
(BRST-closed but not BRST-exact) are the chiral primaries. Finally, this 
anticommutator also implies that any secondary state
 can be decomposed into a 
$\cQ_0$-closed state and a $\cG_0$-closed state. Therefore, we can focus on 
topological descendant states that are either $\cQ_0$-closed or 
$\cG_0$-closed.

 A topological descendant can also be labeled by its level 
$l$ (the $\cL_0$-eigenvalue) and its U(1) charge $(\htop + q)$
(the $\cH_0$-eigenvalue). It is convenient to split the total U(1) charge 
into two pieces: the U(1) charge $\htop$ of the primary state on which the 
descendant is built, thus labeling the corresponding Verma module
 $V(\htop)$, and the
 ``relative" U(1) charge $q$, corresponding to the 
operator acting on the primary field, given by the number of $\cG$ modes 
minus the number of $\cQ$ modes in each term. $\cQ_0$-closed and
$\cG_0$-closed topological secondary states will be
 denoted as $\ket\chi^{(q)Q}$ and 
$\ket\chi^{(q)G}$ respectively (notice that $(q)$ refers to the
relative U(1) charge of the state). 

 It turns out that the topological singular states come
only in four types.
 Namely, $\ket\chi^{(0)G}$, 
$\ket\chi^{(0)Q}$, $\ket\chi^{(1)G}$ and $\ket\chi^{(-1)Q}$
(the fact that these
are the only kinds of topological singular states will be discussed
in \cite{BJI6}). These four 
types can be mapped into each other by using $\cG_0$, $\cQ_0$ 
and the spectral flow automorphism of the
Twisted Topological algebra, denoted by 
$\cA$ \cite{BJI3}. The action of $\cG_0$ and $\cQ_0$ results in the 
following mappings
  
$$\cQ_0\ket\chi_\htop^{(0)G}\rightarrow\ket\chi_\htop^{(-1)Q}, \qquad 
\cG_0\ket\chi_\htop^{(-1)Q}\rightarrow\ket\chi_\htop^{(0)G}, $$
$$\cQ_0\ket\chi_\htop^{(1)G}\rightarrow\ket\chi_\htop^{(0)Q}, \qquad 
\cG_0\ket\chi_\htop^{(0)Q}\rightarrow\ket\chi_\htop^{(1)G}.$$

\noi
Here the subindices indicate the Verma module $V(\htop)$ to which the
states belong. The level of the states
does not change under the action of $\cG_0$ and 
$\cQ_0$, obviously.
It is important to notice that the Verma module does not change 
either, as indicated by the subindices. As a consequence, charged and 
uncharged topological singular
 states come always in pairs in the same Verma module.
 Namely, singular states 
of the types $\ket\chi^{(0)Q}$ and $\ket\chi^{(1)G}$ are together in the 
same Verma module at the same level and a similar statement holds for the 
singular states of the types $\ket\chi^{(0)G}$ and $\ket\chi^{(-1)Q}$.

 The spectral flow automorphism of the
 topological algebra, on the other hand, given by \cite{BJI3}

\BE\new\BA{rclcrcl}
\cA \, \cL_m \, \cA&=& \cL_m - m\cH_m\,,\\
\cA \, \cH_m \, \cA&=&-\cH_m - {\ctop\over3} \delta_{m,0}\,,\\
\cA \, \cQ_m \, \cA&=&\cG_m\,,\\
\cA \, \cG_m \, \cA&=&\cQ_m\,,\
\label{autom} \EA\EE

\noi
changes the Verma module of the states as
 $V(\htop)\rightarrow V(-\htop-{\ctop\over 3}$). 
In addition, $\cA$ reverses 
the relative charge as well as the BRST-invariance properties
 of the states, leaving the 
level invariant. Therefore the action of $\cA$ results in the following 
mappings \cite{BJI3}

\BE
\cA\ket\chi_\htop^{(0)G}\rightarrow\ket\chi_{-\htop-{\ctop\over3}}^{(0)Q}\,
,\qquad 
\cA\ket\chi_\htop^{(-1)Q}\rightarrow\ket\chi_{-\htop-{\ctop\over3}}^{(1)G}\,,
\EE

\noi
with $\cA^{-1} = \cA$.
As a consequence, the topological singular states
come in families of four, one of each kind at the same level.
Two of them, one charged and one uncharged, belong to
the Verma module $V(\htop)$, whereas the other pair belong to a
different Verma module 
 $V(-\htop-{\ctop\over 3})$. This implies that
 there are not ``loose" topological singular states,
\ie\ once one of them exists the other three are generated just by
 the action of \Gn, \Qn\ and $\cA$, as the diagram shows. 

\vskip .2in

\def\xgo  {\mbox{$\kc_{\htop}^{(0)G} $}}
\def\xqo  {\mbox{$\kc_{-\htop-{\ctop\over3}}^{(0)Q} $}}
\def\xqm  {\mbox{$\kc_{\htop}^{(-1)Q} $}}
\def\xgp  {\mbox{$\kc_{-\htop-{\ctop\over3}}^{(1)G} $}}

  \begin{equation}  
  \begin{array}{rcl} \xgo &
  \stackrel{\Qz}{\mbox{------}\!\!\!\longrightarrow}
  & \xqm \\[3 mm]
   \cA\,\updownarrow\ && \ \updownarrow\, \cA
  \\[3 mm]  \xqo \! & \stackrel{\Gz}
  {\mbox{------}\!\!\!\longrightarrow} & \! \xgp  \end{array}
\end{equation}

\vskip .2in
For $\htop=-{\ctop\over6}$ the two Verma modules related by the
spectral flow automorphism coincide.
 Therefore, if there are singular states for this value of $\htop$,
they must come four by four: 
one of each kind at the same level.

To know if there are actually singular states for that or for any other 
value of $\htop$, we have to derive the spectrum of
possible values of $\htop$ corresponding to
the topological Verma modules which contain singular 
states. For this purpose we first have to clarify the relation between the 
topological singular states and the singular states 
 of the untwisted Aperiodic NS algebra.  
After doing this, we will be able to 
exploit the determinant formula 
for the Aperiodic NS algebra.

\section{Untwisting the Topological Singular States}\lvm

The relation between the Twisted Topological algebra
 and the untwisted Aperiodic NS 
 algebra is given, obviously, by the topological twistings
\req{twa} and \req{twb}. These will be called twistings or
untwistings equivalently, depending on the direction of 
the transformation. From the purely
 formal point of view, the Twisted 
Topological algebra \req{topalgebra}
 is simply a rewriting of the untwisted Aperiodic NS
algebra, given by

\BE\new\BA{lclclcl}
\L[L_m,L_n\R]&=&(m-n)L_{m+n}+{\ctop\over12}(m^3-m)\Kr{m+n}
\,,&\qquad&[H_m,H_n]&=
&{\ctop\over3}m\Kr{m+n}\,,\\
\L[L_m,G_r^\pm
\R]&=&\L({m\over2}-r\R)G_{m+r}^\pm
\,,&\qquad&[H_m,G_r^\pm]&=&\pm G_{m+r}^\pm\,,
\\
\L[L_m,H_n\R]&=&{}-nH_{m+n}\\
\L\{G_r^-,G_s^+\R\}&=&\multicolumn{5}{l}{2L_{r+s}-(r-s)H_{r+s}+
{\ctop\over3}(r^2-\frac{1}{4})
\Kr{r+s}\,,}\EA\label{N2algebra}
\EE

\noi
where the fermionic modes take semi-integer values.
The question naturally arises now whether or not the topological singular 
states are also a rewriting of the NS singular states.
 The answer is 
partially affirmative: only \Gn -closed topological singular states
transform into NS singular states after the untwisting. Moreover, these
are built on chiral primaries, when using the twist (2) \req{twb},
or on antichiral primaries, when using the twist (1) \req{twa}.
The twisting of the NS singular states, on the other hand, always
produces \Gn -closed topological singular states. They are built
on topological chiral primaries provided the NS singular states
are built on chiral or antichiral primaries. Otherwise, the 
topological singular states would be built on topological
primaries which are non-chiral and hence not BRST-invariant and
with conformal weights different from zero. Therefore, for the
purpose of twisting and untwisting we are only interested in NS
singular states built on chiral or antichiral primaries.

All these statements can be verified rather easily. One only
needs to investigate how the highest weight (h.w.) conditions
satisfied by the singular states get modified under the
twistings or untwistings given by \req{twa} and \req{twb}.
By inspecting these, it is obvious that the bosonic h.w.
conditions, \ie\ $L_{m>0}\ket{\chi_{NS}}=H_{m>0}\ket{\chi_{NS}}=0$
on the one hand, and $\cL_{m>0}\ket\chi=\cH_{m>0}\ket\chi=0$
on the other hand, are conserved under the twistings and untwistings.
In other words, if the topological state
 $\ket\chi$ satisfies the topological
bosonic h.w. conditions, then the corresponding untwisted state
$\ket{\chi_{NS}}$ satisfies the NS bosonic h.w. conditions
and vice versa. With the fermionic h.w. conditions things are
not so straightforward. While the h.w. conditions 
$\cQ_{m>0}\ket\chi=0$ are converted into h.w. conditions of the
type $G_{m\geq\half}^{\pm}\ket{\chi_{NS}}=0$
($G^+$ or $G^-$ depending on the specific twist), in both twistings
one of the $G_{1/2}^{\pm}$ modes is transformed into \Gn\ . But
$G_{1/2}^{\pm}\ket{\chi_{NS}}=0$ is nothing but a h.w. condition satisfied 
by all the NS singular states!

As a result, the twisting of a NS singular state (using
 \req{twa} if it is built on an antichiral primary, or 
 \req{twb} if it is built on a chiral primary), always
produces a \Gn -closed topological singular state. Conversely,
only \Gn -closed topological singular states, \ie\ those of the
types $\kc^{(0)G}$ and $\kc^{(1)G}$, produce NS singular states
under the untwistings.

Now let us analyze the transformation of the $(\cL_0, \cH_0)$
eigenvalues $(l, q+ \htop )$ of the topological singular states
$\kc_l^{(q)G}$ into $(L_0, H_0)$ eigenvalues
$(\Delta + l, q+ \htop )$ of the untwisted NS singular states
$\ket{\chi_{NS}}_l^{(q)}$. In other words, given a topological
 singular state in the Verma module $V(\htop)$, at level
$l$ and with relative charge $q$, let us determine the Verma 
module, the level and the relative charge of the corresponding
untwisted NS singular state.

Using the twist (1) \req{twa} the U(1) charge does not change at
all, while the level gets modified as $l- \half q$. Therefore
the topological singular states of the types $\kc_l^{(0)G}$
and $\kc_l^{(1)G}$ are transformed into NS singular states
of the types $\ket{\chi_{NS}}_l^{(0)a}$ and
$\ket{\chi_{NS}}_{l-\half}^{(1)a}$ built on antichiral primaries,
as is indicated by the superscript $a$.
The Verma modules which contain the topological singular states, 
say $V (\htop^{(0)})$ for the uncharged states $\kc_l^{(0)G}$
and $V (\htop^{(1)})$ for the charged states $\kc_l^{(1)G}$
(with the relation $\htop^{(1)} = - \htop^{(0)} - {\ctop \over 3}$
imposed by the spectral flow automorphism of the Topological
algebra), transform simply as $V_{NS} (\htop^{(0)})$ and
$V_{NS} (\htop^{(1)})$. That is, using the twist (1), the topological
 Verma modules
transform into NS Verma modules, but the U(1)
charge of the primary states on which they are built remains
unmodified. The conformal weights of the antichiral primary states
are related to their U(1) charges as
 $\Delta = -{\htop \over 2}$, as is well known.

Using the twist (2) \req{twb}, on the other hand, the U(1)
charge reverses its sign, while the level gets modified, again
as $l - \half q$. Therefore the topological singular states of
the types $\kc_l^{(0)G}$
and $\kc_l^{(1)G}$ result in NS singular states
of the types $\ket{\chi_{NS}}_l^{(0)ch}$ and
$\ket{\chi_{NS}}_{l-\half}^{(-1)ch}$ built on chiral primaries,
as is indicated by the superscript $ch$.
The corresponding Verma modules $V (\htop^{(0)})$ and
$V (\htop^{(1)})$ are transformed into the NS Verma modules
$V_{NS} (-\htop^{(0)})$ and $V_{NS} (-\htop^{(1)})$. The conformal
weights of the chiral primary states, in turn,  are related to their
U(1) charges as $\Delta = {\htop \over 2}$.

An interesting observation here is that the same type of topological
singular state $\kc_l^{(1)G}$ gives rise to both the charge $(+1)$
and the charge $(-1)$ NS singular states at level $l-\half$, built
on antichiral primaries and chiral primaries respectively.
Furthermore, since there are no topological singular states of
the type $\kc_l^{(-1)G}$, we deduce that {\it all} the charged
NS singular states built on chiral primaries have relative 
U(1) charge $(-1)$, whereas {\it all} the charged NS singular
states built on antichiral primaries have relative U(1)
charge $(+1)$. Moreover, the charge $(+1)$ and charge $(-1)$
singular states are mirrored under the interchange
$H_m \leftrightarrow - H_m$ (therefore $\htop \leftrightarrow
-\htop$) and $G_r^+ \leftrightarrow G_r^-$.

Finally, let us stress the fact that the charged and uncharged
NS singular states $\ket{\chi_{NS}}_{l}^{(0)a}$
and $\ket{\chi_{NS}}_{l-\half}^{(1)a}$, on the one hand, and
$\ket{\chi_{NS}}_{l}^{(0)ch}$
and $\ket{\chi_{NS}}_{l-\half}^{(-1)ch}$, on the other hand,
 must come in pairs, although
in different Verma modules, since they are just the untwistings
of the topological singular states $\kc_l^{(0)G}$ and $\kc_l^{(1)G}$,
which come in pairs inside the four-member topological families
described in last section.

\section{Spectrum of Topological and NS Singular States}\lvm

As we have just shown, the spectrum of U(1) charges corresponding
to the Verma modules which contain
topological and NS singular states
is the same for the singular states of the
types $\kc_l^{(0)G}$, $\kc_l^{(-1)Q}$ and $\ket{\chi_{NS}}_l^{(0)a}$, 
and also the same 
 for the singular states of the
types $\kc_l^{(0)Q}$, $\kc_l^{(1)G}$ and
 $\ket{\chi_{NS}}_{l-\half}^{(1)a}$, where the superscript $a$
indicates that the NS singular states are built on antichiral
primaries. The first spectrum is given by all the possible values of
 $\htop^{(0)}$ whereas the second one is given by all the
 possible values of $\htop^{(1)}$, these values being connected
to each other by the relation  $\htop^{(1)} =
-\htop^{(0)} - {\ctop\over3}$.

The spectrum for the case of NS Verma modules  built on chiral
 primaries, in turn, is the corresponding to $(-\htop^{(0)})$,
for singular states of the type $\ket{\chi_{NS}}_l^{(0)ch}$,
and to $(-\htop^{(1)})$, for singular
states of the type $\ket{\chi_{NS}}_{l-\half}^{(-1)ch}$.

In principle, we cannot compute the spectra $\htop^{(1)}$ and
 $\htop^{(0)}$ of charged and uncharged singular states 
built on antichiral primaries, simply by imposing 
the relation $\Delta = - {\htop \over2}$ in the spectra given
 by the determinant formula of the Aperiodic NS algebra 
\cite{BFK}. The reason is that those spectra do not apply for
 incomplete Verma modules constructed on chiral or antichiral
 primary states, \ie\ highest weight states $\ket{\Delta,\htop}$
 which are annihilated by $G_{-\half}^+$ or $G_{-\half}^-$
 (for which, as a result, $\Delta ={\htop \over2}$ or 
$\Delta = - {\htop \over2}$ respectively).
Hence, it seems that one has to
 compute a new determinant formula specific for 
chiral representations. However, this is not necessary since,
as we will see, the zeroes of the determinant formula contain
the zeroes of the ``chiral" determinant formula in a simple,
easy to identify, pattern.

We start our analysis with the ansatz
that the set of zeroes of the chiral determinant formula 
 is included in the set of zeroes of the general determinant
 formula for the particular cases $\Delta ={\htop \over2}$
(or $\Delta =-{\htop \over2}$), with possibly a different 
interpretation in terms of h.w. singular states. Although this ansatz
seems rather intuitive, the proof is not straightforward at all.
 The reason is that the equations obtained by imposing
 the h.w. conditions on a given secondary state
 depend on whether the primary state is chiral or not. 
As a consequence, the h.w. conditions give
 different h.w. singular states for chiral representations
 than for non-chiral representations (see the Appendix).

 Our strategy will be now to analyze the zeroes of the determinant formula 
taking into account that the spectral flow automorphism of the topological 
algebra predicts an equal number of charged and uncharged singular states 
(built on chiral or antichiral primaries) with a precise relation between 
their corresponding Verma modules. Namely, for singular states on antichiral 
modules the relation is $\htop^{(1)}=-\htop^{(0)}-{\ctop\over3}$, while for 
singular states on chiral modules it is
 $\htop^{(1)}=-\htop^{(0)}+{\ctop\over3}$
 (in this last expression $\htop^{(1)}$ 
and $\htop^{(0)}$ denote the spectra 
 for the chiral modules, instead of ($-\htop^{(1)}$) and 
($-\htop^{(0)}$) in our previous notation).

Let us concentrate on the antichiral modules, for convenience. The zeroes of 
the determinant formula of the Aperiodic NS algebra \cite{BFK}
are given by the solutions of the quadratic 
vanishing surface $f_{rs}^A=0$, with

\BE   f_{r,s}^A = 2 \L({\ctop-3\over3}\R) \Delta - \htop^2
 -{1\over4} \L({\ctop-3\over3}\R)^2 +
 {1\over4} \L(\L({\ctop-3\over3}\R) r + s \R)^2 \qquad r\in\oZ^+\,,\,\,
  s\in2\oZ^+ \label{frs} \EE

\noi
and the solutions of the vanishing plane $g_k^A = 0$,
with
       
\BE   g_k^A = 
  2 \Delta-2k\htop + \L({\ctop-3\over3}\R)(k^2-{1\over4})
\qquad k\in\oZ+\half \label{gk} \EE

\noi
Solving for $f_{r,s}^A = 0$, with $\Delta = - {\htop\over2}$, one finds
two two-parameter solutions for $\htop$ (since
$f_{r,s}^A = 0$ becomes a quadratic equation for $\htop$).
These solutions are

\BE  \htop_{r,s} = - \half \L(
      \L({\ctop-3\over3}\R) (r+1) + s \R)  \label{h0rs} \EE
\noi 
and

\BE  \hat\htop_{r,s} = \half \L(
      \L({\ctop-3\over3}\R) (r-1) + s \R)\,.  \label{hh0rs} \EE

\noi                                                   
Solving for $g_k^A = 0$, with $\Delta = - {\htop\over2}$,
 one finds the one-parameter solution

\BE \htop_k =  \L({\ctop-3\over6}\R) (k-\half)\,,
                                          \label{hk}\EE

\noi
except for $k=-\half$ where $g_k^A$ is identically zero.

There are two different situations to be considered now, 
depending on whether the number of zeroes of the chiral
determinant formula is equal or smaller than the number of zeroes 
given by the solutions \req{h0rs}, \req{hh0rs} and \req{hk}.

Let us start with the first situation. The problem at hand is
therefore to distribute all these solutions into two sets,
say $H^{(0)}$ and $H^{(1)}$, such that for any given
solution $\htop^{(0)}$ in the set $H^{(0)}$
there exists one solution $\htop^{(1)}$ in the set $H^{(1)}$,
satisfying $\htop^{(1)} = -\htop^{(0)} - {\ctop\over3}$,
and vice versa. For this purpose it is helpful to write
down the set of expressions corresponding to
$(-\htop_{r,s} - {\ctop\over3})$,
 ${\ }(-\hat\htop_{r,s} - {\ctop\over3}){\ }$
and ${\ }(-\htop_k - {\ctop\over3})$. These are given by

\BE -\htop_{r,s} - {\ctop\over3} {\ } = {\ } \half \L(
      \L({\ctop-3\over3}\R) (r-1) + s - 2 \R)\,,  \label{mh0rs} \EE

\BE -\hat\htop_{r,s} - {\ctop\over3} {\ } = - \half \L(
      \L({\ctop-3\over3}\R) (r+1) + s + 2 \R)  \label{mhh0rs} \EE

\noi
and

\BE -\htop_k - {\ctop\over3} {\ } = - \half \L(
      \L({\ctop-3\over3}\R) (k+{3\over2}) + 2 \R)\,.  \label{mhk} \EE

Comparing these expressions with the set of solutions
$\htop_{r,s}{\ }$, $\hat\htop_{r,s}{\ }$ and ${\ }\htop_k{\ }$, given by
\req{h0rs}, \req{hh0rs} and \req{hk}, one finds straightforwardly

\BE   \htop_{r,2} = - \htop_{r-\half} - {\ctop\over3}\,, \qquad
    \htop_{r,s>2} = - \hat\htop_{r,(s-2)} - {\ctop\over3}\,.
\EE

\noi
Therefore one solution to the problem is that the
spectrum of the uncharged singular states is given by
${\ }\htop_{r,s}^{(0)} = \htop_{r,s}{\ }$, with the level of the
state $l = {rs\over2}$, whereas the spectrum of the charged
singular states, at level $l-\half={rs-1\over2}$, 
 is given by 
${\ }\htop_{r,s}^{(1)} = - \htop_{r,s} - {\ctop\over3}{\ }$, that is

\BE  \htop_{r,s}^{(1)} = \half \L(
      \L({\ctop-3\over3}\R) (r-1) + s - 2 \R)\,,  \label{h1rs} \EE
 
\noi
which contains the two series  $\htop_k$ and
$\hat\htop_{r,s}{\ }$. Namely, for $s=2{\ }$
$\htop_{l,2}^{(1)} = \htop_{l-\half}$, where $l-\half$ is
the level of the charged singular state, while for $s>2{\ }$
$\htop_{r,s>2}^{(1)} = \hat\htop_{r,(s-2)}$, with the level
$l-\half$ given by $l-\half = {rs-1 \over2}{\ }$.

We see therefore that in this solution half of the zeroes
of the quadratic vanishing surface $f_{r,s}^A = 0$ 
correspond to uncharged singular states at level $l={rs\over2}$,
and the other half correspond to charged singular states at
level $l-\half = {r(s+2)-1 \over2}{\ }$. The zeroes of the 
vanishing plane $g_k^A = 0$, in turn, correspond to charge $(+1)$
singular states at level $k$, built on antichiral primaries,
for $k>0$, and to charge $(-1)$
singular states at level $(-k)$, built on chiral primaries,
for $k<0$ (the reader can verify this last statement by 
repeating our analysis imposing chirality rather than
antichirality on the primary states).

The second situation, in which the number of zeroes of the
chiral determinant formula is smaller than the number of
zeroes of the general determinant formula has, in principle, more
solutions than the previous one. The simplest one is to set
  $ \htop^{(0)} = \htop_{l,2} {\ }, {\ }
 \htop^{(1)} = -\htop_{l,2} - {\ctop\over3} = \htop_{l-\half}{\ }$.
However, before searching for more intricate possibilities
let us have a look at the data for $\htop^{(0)}$ and
$\htop^{(1)}$ given by the singular states
themselves. We have computed until level 3, just by imposing the
h.w. conditions, all the topological singular states and all
the NS singular states built on chiral and antichiral primaries
(some of these singular states were already published, of course). 
The explicit expressions for all these singular states will be
given in \cite{BJI6}, although in the Appendix we also write down
and analyze the NS singular states $\ket{\chi_{NS}}_{1}^{(0)}$, 
$\ket{\chi_{NS}}_{3\over2}^{(1)a}$ and
 $\ket{\chi_{NS}}_{3\over2}^{(-1)ch}$. 
 Here we need only the values of $\htop^{(0)}$ and $\htop^{(1)}$ .
These are the following:
\vskip .2in

- For $\ket\chi_1^{(0)G},{\ } \ket\chi_1^{(-1)Q}$  and 
$\ket{\chi_{NS}}_1^{(0)a} {\ \ } \htop^{(0)}= -{\ctop\over3}$

\vskip .17in

- For $\ket\chi_1^{(0)Q},{\ } \ket\chi_1^{(1)G}{\ }$ and 
$\ket{\chi_{NS}}_{\half}^{(1)a}{\ \ } \htop^{(1)}= 0$ 

\vskip .17in

- For $\ket\chi_2^{(0)G},{\ } \ket\chi_2^{(-1)Q}$ and 
$\ket{\chi_{NS}}_2^{(0)a} {\ \ } \htop^{(0)}=
 {1-\ctop\over2}{\ },{\ \ } -{\ctop+3\over3} $

\vskip .17in

- For $\ket\chi_2^{(0)Q},{\ } \ket\chi_2^{(1)G}{\ }$ and 
$\ket{\chi_{NS}}_{3\over2}^{(1)a}{\ \ } \htop^{(1)}={\ctop-3\over 6}{\ }, 
{\ \ }1$                                                 

\vskip .17in

- For $\ket\chi_3^{(0)G},{\ }\ket\chi_3^{(-1)Q}$ and 
$\ket{\chi_{NS}}_3^{(0)a}{\ \ }\htop^{(0)}={3-2\ctop\over3}{\ },{\ \ }
-{\ctop+6\over3}$

\vskip .17in

- For $\ket\chi_3^{(0)Q},{\ }\ket\chi_3^{(1)G}{\ }$ and 
$\ket{\chi_{NS}}_{5\over2}^{(1)a}{\ \ }\htop^{(1)}={\ctop-3\over3}{\ },
{\ \ }2\,.$

\vskip .2in

In addition, the BRST-invariant uncharged topological
singular state at level 4 $\ket\chi_4^{(0)Q}$ has been
computed in \cite{Sem} with the result
${\ }\htop^{(1)} = {\ctop-3\over2}, {\ }{\ctop+3\over6}, {\ }3$.
These values also correspond to the singular states 
${\ }\ket\chi_4^{(1)G}{\ }$ and 
$\ket{\chi_{NS}}_{7\over2}^{(1)a}$, as we have deduced in section 3.

By comparing these results with the zeroes of the
determinant formula, given by
expressions \req{h0rs}, \req{hh0rs} and \req{hk},
 we notice that the values
we have found for $\htop^{(0)}$ fit exactly in the 
expression $\htop_{r,s}$ \req{h0rs}, that is, the upper
solution for the quadratic vanishing surface,
but not in the lower solution
$\hat\htop_{r,s}$. 
  The values we
have found for $\htop^{(1)}$ follow exactly the prediction of
the spectral flow automorphism for each case, \ie\
${\ }\htop^{(1)} = -\htop^{(0)} - {\ctop\over3}{\ }$.
Therefore we can set
 $\htop_{r,s}^{(0)} = \htop_{r,s}$ and  
 $\htop_{r,s}^{(1)} = - \htop_{r,s} - {\ctop\over3}$.            
Hence, the actual spectra of singular states follow exactly
the pattern we have found under the ansatz that the zeroes
of the chiral determinant formula coincide with the zeroes
of the determinant formula (for $\Delta = -{\htop\over2}$
in the antichiral case). That is, $\htop_{r,s}^{(0)}$ is
given by \req{h0rs} and $\htop_{r,s}^{(1)}$ is given
by \req{h1rs}.

For the case of NS singular states built on chiral primaries 
one finds the same values for the U(1) charges as for the
antichiral case, but with the sign
reversed, as expected.
 Therefore the spectra of U(1) charges for the
NS singular states of the types $\ket{\chi_{NS}}_{l}^{(0)ch}$
and $\ket{\chi_{NS}}_{l-\half}^{(-1)ch}$ are given by 
$(-\htop^{(0)}_{r,s})$ and $(-\htop^{(1)}_{r,s})$ respectively.

\vskip .2in

Let us summarize our results for the spectra of U(1) charges
and conformal weights corresponding to the Verma modules
 which contain the different kinds of singular states.

\vskip .17in

*) The spectrum of U(1) charges corresponding to topological
singular states of the types $\kc_l^{(0)G}$ and $\kc_l^{(-1)Q}$,
and uncharged NS singular states built on antichiral primaries
$\ket{\chi_{NS}}_{l}^{(0)a}$ is given by ${\ }\htop^{(0)}_{r,s}{\ }$,
eq.\req{h0rs}, where the level of the singular states is            
 given by $l = {r s \over 2}$. The spectrum of conformal
weights for the NS singular states $\ket{\chi_{NS}}_{l}^{(0)a}$
is given therefore by
 $\Delta^{(0)}_{r,s} = -\half \htop^{(0)}_{r,s}$.

\vskip .17in

*) The spectra of U(1) charges and conformal weights 
corresponding to uncharged NS singular states built on chiral primaries
$\ket{\chi_{NS}}_{l}^{(0)ch}$ are given by $(-\htop^{(0)}_{r,s})$  
and $\Delta^{(0)}_{r,s} = \half (-\htop^{(0)}_{r,s})$. 
Therefore the spectrum of conformal weights corresponding to
uncharged NS singular states is the same for those built on
chiral primaries than for those built on antichiral primaries,
although the spectrum of U(1) charges is reversed in sign.

\vskip .17in

*) The spectrum of U(1) charges corresponding to topological
singular states of the types $\kc_l^{(0)Q}$ and $\kc_l^{(1)G}$,
and charged NS singular states built on antichiral primaries
$\ket{\chi_{NS}}_{l-\half}^{(1)a}$ is given by
 ${\ }\htop^{(1)}_{r,s} = -\htop^{(0)}_{r,s} - {\ctop\over3}{\ }$,
resulting in expression \req{h1rs},
 where $l = {r s \over 2}$. The spectrum of conformal weights
 for the NS singular states $\ket{\chi_{NS}}_{l-\half}^{(1)a}$
is given therefore by
 $\Delta^{(1)}_{r,s} = -\half \htop^{(1)}_{r,s}$.
 
\vskip .17in

*) The spectra of U(1) charges and conformal weights 
corresponding to charged NS singular states built on chiral primaries
$\ket{\chi_{NS}}_{l-\half}^{(-1)ch}$ are given by $(-\htop^{(1)}_{r,s})$  
and $\Delta^{(1)}_{r,s} = \half (-\htop^{(1)}_{r,s})$. 
Therefore the spectrum of conformal weights corresponding to
charged NS singular states is the same for those built on
chiral primaries than for those built on antichiral primaries,
although the spectrum of U(1) charges, as well
as the relative charge, is reversed in sign.

\vskip .2in
Comparing these spectra with the BFK spectra
for non-chiral representations, one
finds the remarkable fact that half of the uncharged singular
states with levels $l={rs\over2}$, in the non-chiral Verma modules,
 have been replaced by charged singular states with levels
$l-\half = {r(s+2)-1\over2}$, in the chiral modules. Namely those singular 
states with $\htop$ given by $\pm \hat\htop_{rs}$ in
 \req{hh0rs}. In the Appendix we
analyze this effect for the particular case of the uncharged
singular states  $\ket{\chi_{NS}}_1^{(0)}$ and the
charged singular states 
$\ket{\chi_{NS}}_{3\over2}^{(1)}$ and
 $\ket{\chi_{NS}}_{3\over2}^{(-1)}$.
We write down the h.w. equations, with their solutions, for
the primary states being non-chiral, chiral and antichiral.

An interesting question now is which kind of states become the
uncharged singular states with $\htop=\pm \hat\htop_{rs}$
once we switch on chirality on the primary states. An equally
interesting question is which kind of states are the charged singular
states with $\htop= \pm \hat\htop_{rs}$ once we switch off
chirality on the primary states. We do not know the answer in general.
 In the Appendix we show that the uncharged singular
states $\ket{\chi_{NS}}_1^{(0)}$, for $\Delta = \mp {\htop\over2}{\ },$
 $\htop = \pm \hat\htop_{1,2}= \pm 1{\ }$ vanish once we switch on
antichirality and chirality on the primary states, respectively.
 The charged singular states 
$\ket{\chi_{NS}}_{3\over2}^{(1)a}$ with $\htop=1$ and 
 $\ket{\chi_{NS}}_{3\over2}^{(-1)ch}$ with $\htop=-1$, 
 on the other hand, become non-highest
weight singular states after we switch off antichirality and chirality
respectively on the primary states. These singular states are very
special since, in spite of being non-highest weight states, they are not
secondary states of any h.w. singular states either (they can descend
down to a h.w. singular state but not the other way around).

\section{Spectrum of R Singular States}\lvm

The singular states of the Aperiodic NS algebra transform into
singular states of the Periodic R algebra under the action of
the spectral flows, and vice versa \cite{SS}, \cite{Kir1},
 \cite{LVW}, \cite{BJI4}. In particular, the NS singular states built
on chiral or antichiral primaries transform into R singular states
built on the Ramond ground states. As a consequence, we can
write down easily the spectrum of U(1) charges for the
 R singular states built on
the Ramond ground states simply by applying the spectral flow
transformations to the spectra \req{h0rs} and \req{h1rs}
found in last section. Before doing this let us say a few
words about the Periodic R algebra.

The Periodic N=2 Superconformal algebra is given by
\req{N2algebra}, where the fermionic generators
$G^{\pm}_r$ are integer moded. The zero modes of the
fermionic generators characterize the states as being
$G_0^+$-closed or $G_0^-$-closed, as the anticommutator
${\ }\{G_0^+,G_0^-\} = 2L_0 - {\ctop\over12}{\ }$ shows.
The Ramond ground states are annihilated by both $G_0^+$
and $G_0^-$, therefore $\Delta = {\ctop\over24}$ for
them, as a result.

In order to simplify the analysis that follows it is very
convenient to define the U(1) charge for the states of the
Periodic R algebra in the same way as for the states of the
Aperiodic NS algebra. Namely, the U(1) charge of the states
will be denoted by $\htop$, instead of $\htop \pm \half$,
whereas the relative charge $q$ of a secondary state will
be defined as the difference between the $H_0$-eigenvalue
of the state and the $H_0$-eigenvalue of the primary on which it
is built. Therefore, the relative charges of the R states
are defined to be integer, in contrast with the usual definition.

The spectral flow \cite{SS}, \cite{LVW}, is given by the
one-parameter family of transformations

\BE\new\BA{rclcrcl}
\cU_\th \, L_m \, \cU_\th^{-1}&=& L_m
 +\th H_m + {\ctop\over 6} \th^2 \delta_{m,0}\,,\\
\cU_\th \, H_m \, \cU_\th^{-1}&=&H_m + {\ctop\over3} \th \delta_{m,0}\,,\\
\cU_\th \, G^+_r \, \cU_\th^{-1}&=&G_{r+\th}^+\,,\\
\cU_\th \, G^-_r \, \cU_\th^{-1}&=&G_{r-\th}^-\,,\
\label{spfl} \EA\EE

\noi
satisfying $\cU_{\th}^{-1} = \cU_{(-\th)}$ and giving rise to
isomorphic algebras. If we denote by $(\Delta, \htop)$ the 
$(L_0, H_0)$ eigenvalues of any given state, then 
the eigenvalues of the transformed state
$\cU_{\th} \kc$ are
 $(\Delta-\th \htop +{c\over6} \th^2, \htop - {c\over3} \th)$.
If the state $\kc$ is a level-$l$ secondary state with relative
 charge $q$ and eigenvalues $(\Delta+l, \htop+q)$ (where now
$(\Delta, \htop)$ denote the eigenvalues of the primary on
which the secondary is built), then one gets straightforwardly
that the level of the transformed state $\cU_{\th} \kc$ changes
to $l-\th q$, while the relative charge remains equal. 

There is another spectral flow \cite{BJI4}, which is the
untwisting of the topological algebra automorphism, given by

\BE\new\BA{rclcrcl}
\cA_\th \, L_m \, \cA_\th&=& L_m
 +\th H_m + {\ctop\over 6} \th^2 \delta_{m,0}\,,\\
\cA_\th \, H_m \, \cA_\th&=&- H_m - {\ctop\over3} \th \delta_{m,0}\,,\\
\cA_\th \, G^+_r \, \cA_\th&=&G_{r-\th}^-\,,\\
\cA_\th \, G^-_r \, \cA_\th&=&G_{r+\th}^+\,.\
\label{ospfl} \EA\EE

\noi
with $\cA_{\th}^{-1} = \cA_{\th}$.
The $(L_0, H_0)$ eigenvalues of the transformed states 
$\cA_{\th} \kc$ are now
 $(\Delta+\th \htop +{c\over6} \th^2, - \htop - {c\over3} \th)$
(that is, they differ from the previous case by the
 interchange $\htop \rightarrow -\htop$). From this one easily
deduces that, under the spectral flow \req{ospfl}, the level $l$
of any descendant will change to $l + \th q$ while  
the relative charge $q$ reverse its sign.

For half-integer values of $\theta$ the two spectral flows
interpolate between the Aperiodic NS algebra and the Periodic
R algebra. In particular, for $\theta = \half$ the primaries
of the NS algebra (including singular states) become primaries
of the R algebra with chirality $(-)$ (\ie\ annihilated by
$G_0^-$), while for $\theta = -\half$ the primaries of the NS
algebra become primaries of the R algebra with chirality
 (+) (\ie\ annihilated by $G_0^+$). In addition, 
$\cU_{1/2}$ and $\cA_{-1/2}$ map the chiral primaries of
the NS algebra (\ie\ annihilated by $G^+_{-1/2}$) into
the set of Ramond ground states, whereas
$\cU_{-1/2}$ and $\cA_{1/2}$ map the antichiral primaries
(\ie\ annihilated by $G^-_{-1/2}$) into the set of
Ramond ground states. As a result, the spectral flows
\req{spfl} and  \req{ospfl}
 transform the NS singular states 
built on chiral and antichiral primaries into R singular states
built on the Ramond ground states, as we said before.
We will denote the R states as $\ket{\chi_R}_l^{(q)+}$ and
$\ket{\chi_R}_l^{(q)-}$, where, in addition to the level and the 
relative charge, we indicate that the state is annihilated by
$G_0^+$ or $G_0^-$.  

As we showed in section 3, the NS singular states built on chiral
 primaries are only of two types, $\ket{\chi_{NS}}_l^{(0)ch}$
and $\ket{\chi_{NS}}_{l-\half}^{(-1)ch}$, and similarly, the
NS singular states built on antichiral primaries come only
in two types $\ket{\chi_{NS}}_l^{(0)a}$ and 
$\ket{\chi_{NS}}_{l-\half}^{(1)a}$. In fact, $\ket{\chi_{NS}}_l^{(0)ch}$
and $\ket{\chi_{NS}}_l^{(0)a}$, on the one hand, and
$\ket{\chi_{NS}}_{l-\half}^{(-1)ch}$ and $\ket{\chi_{NS}}_{l-\half}^{(1)a}$,
on the other hand, are mirrored under the interchange
 $H_m \leftrightarrow -H_m$, $G_r^+ \leftrightarrow G_r^-$, because
they are the two possible untwistings of the \Gn -closed
topological singular states $\kc^{(0)G}_l$ and $\kc^{(1)G}_l$
respectively. 

{}From the discussion above one deduces easily that the NS singular
states $\ket{\chi_{NS}}_l^{(0)a}$, $\ket{\chi_{NS}}_{l-\half}^{(1)a}$,
$\ket{\chi_{NS}}_l^{(0)ch}$ and 
$\ket{\chi_{NS}}_{l-\half}^{(-1)ch}$ are transformed into R singular
states in the following way:

\BE \cA_{1/2} {\ } \ket{\chi_{NS}}_l^{(0)a} = 
\cU_{1/2} {\ } \ket{\chi_{NS}}_l^{(0)ch}=\ket{\chi_R}_l^{(0)-} \EE

\BE \cA_{1/2} {\ } \ket{\chi_{NS}}_{l-\half}^{(1)a} = 
\cU_{1/2} {\ } \ket{\chi_{NS}}_{l-\half}^{(-1)ch}=\ket{\chi_R}_l^{(-1)-}\EE

\BE \cU_{-1/2} {\ } \ket{\chi_{NS}}_l^{(0)a} = 
\cA_{-1/2} {\ } \ket{\chi_{NS}}_l^{(0)ch}=\ket{\chi_R}_l^{(0)+} \EE

\BE \cU_{-1/2} {\ } \ket{\chi_{NS}}_{l-\half}^{(1)a} = 
\cA_{-1/2} {\ } \ket{\chi_{NS}}_{l-\half}^{(-1)ch}=\ket{\chi_R}_l^{(1)+}\EE

\noi
where the NS Verma modules $V_{NS}(\htop)$ are transformed, in turn, as 

\vskip .17in
${\ }{\ }{\ }$
${\ }{\ }{\ } \cU_{1/2}{\ } V_{NS}(\htop)
 \rightarrow V_R(\htop-{\ctop\over6}){\ }$, 
${\ }{\ }{\ }\cU_{-1/2}{\ }V_{NS}(\htop)
 \rightarrow V_R(\htop+{\ctop\over6}){\ }$,\\
${\ }{\ }{\ }{\ }$
${\ }{\ }{\ }\cA_{1/2}{\ } V_{NS}(\htop)
 \rightarrow V_R(-\htop-{\ctop\over6}){\ }$,
${\ }{\ }{\ }\cA_{-1/2}{\ }V_{NS}(\htop)
 \rightarrow V_R(-\htop+{\ctop\over6}){\ }$.

\vskip .17in
Observe that the charged singular states built on the Ramond ground
states come only in two types: $\ket{\chi_R}_l^{(1)+}$ and
$\ket{\chi_R}_l^{(-1)-}$, \ie\ the sign of the relative charge is
equal to the sign of the chirality.

Hence, whereas the spectrum of conformal weights for all these R
singular states is just given by $\Delta={\ctop\over24}$, the 
spectrum of U(1) charges is given by: 
${\ }\htop_{r,s}^{(0)+} = \htop_{r,s}^{(0)} + {\ctop\over6}{\ }$
for singular states of type $\ket{\chi_R}_l^{(0)+}$,
${\ }\htop_{r,s}^{(1)+} = \htop_{r,s}^{(1)} + {\ctop\over6}{\ }$
for singular states of type $\ket{\chi_R}_l^{(1)+}$,
${\ }\htop_{r,s}^{(0)-} = - (\htop_{r,s}^{(0)} + {\ctop\over6}){\ }$
for singular states of type $\ket{\chi_R}_l^{(0)-}$
and ${\ }\htop_{r,s}^{(-1)-}=-(\htop_{r,s}^{(1)} + {\ctop\over6}){\ }$
for singular states of type $\ket{\chi_R}_l^{(-1)-}$.

Using the expressions for ${\ }\htop_{r,s}^{(0)}{\ }$
 and ${\ }\htop_{r,s}^{(1)}{\ }$
given by \req{h0rs} and \req{h1rs} we obtain
finally the spectra
of U(1) charges for the R singular states built on the Ramond
ground states:

\BE \htop_{r,s}^{(0)+} = -\half \L( \L({\ctop-3\over3} \R)r+s-1 \R)
 {\ } {\rm for} \qquad \ket{\chi_R}_l^{(0)+}{\ }, \label{h0+rs} \EE

\BE \htop_{r,s}^{(0)-} = \half \L( \L({\ctop-3\over3} \R)r+s-1 \R)
 {\ } {\rm for} \qquad \ket{\chi_R}_l^{(0)-}{\ }, \label{h0-rs} \EE

\BE \htop_{r,s}^{(1)+} = \half \L( \L({\ctop-3\over3} \R)r+s-1 \R)
 {\ } {\rm for} \qquad \ket{\chi_R}_l^{(1)+}{\ }{\ } \label{h1+rs} \EE
\noi
and

\BE \htop_{r,s}^{(-1)-} = -\half \L( \L({\ctop-3\over3} \R)r+s-1 \R)
 {\ } {\rm for} \qquad \ket{\chi_R}_l^{(-1)-}{\ }.\label{h1-rs} \EE

We see that $\ket{\chi_R}_l^{(0)+}{\ }$ and $\ket{\chi_R}_l^{(-1)-}{\ }$
are together in the same Verma module $V_R(\htop)$ and at the same level,
and so are $\ket{\chi_R}_l^{(0)-}{\ }$ and $\ket{\chi_R}_l^{(1)+}{\ }$,
which belong to the Verma module $V_R(-\htop)$. Therefore, the 
singular states built on the Ramond ground states come always
in sets of two pairs at the same level. Every pair consists of
one charged and one uncharged state, with opposite chiralities,
one pair belonging to the Verma module $V_R(\htop)$ and the
other to the Verma module $V_R(-\htop)$. This is easy to see also
taking into account that the action of $G_0^+$ or $G_0^-$ on
any singular state built on the Ramond ground states produces
another singular state with different relative charge and different
chirality but with the same level and sitting in the same Verma
module (since both $G_0^+$ and $G_0^-$ annihilate the Ramond
ground states).

This resembles very much the family structure for the Twisted
 Topological algebra that we analyzed in section 2. However, 
there is a drastic difference here because in the latter case the
four members of the topological family, \ie\ 
$\kc_l^{(0)G}$, $\kc_l^{(0)Q}$, $\kc_l^{(1)G}$ and $\kc_l^{(-1)Q}$,
are completely different from each other, while in this case the four
 members are two by two mirror symmetric under the interchange 
$H_m \leftrightarrow -H_m{\ }$, $G_r^+ \leftrightarrow G_r^-$.

Now let us compare the spectra we have found, eqns. \req{h0+rs},
\req{h0-rs}, \req{h1+rs} and \req{h1-rs}, with the zeroes of the
determinant formula for the Periodic R algebra 
for $\Delta = {\ctop\over 24}$. Let us remind that our
definition of U(1) charge is different from the definition given
in \cite{BFK}. For example, our uncharged states are called
 charge ($-\half sgn(0))$ states there.

The zeroes of the determinant formula for the Periodic R algebra
are given by the vanishing quadratic surface $f_{r,s}^P=0$
and the vanishing plane $g_k^P=0$, where

\BE f_{r,s}^P = 2\L({\ctop-3\over3}\R)(\Delta-{\ctop\over24}) - \htop^2
+{1\over4}\L(\L({\ctop-3\over3}\R) r+s \R)^2 \qquad r\in\oZ^+\,,\,\,
  s\in2\oZ^+ \label{frsP} \EE

\noi
and 
       
\BE   g_k^P = 
  2 \Delta-2k\htop + \L({\ctop-3\over3}\R)(k^2-{1\over4})
 - {1\over4} \qquad k\in\oZ+\half \,. \EE

\noi
For $\Delta={\ctop\over24}$ they result in the following solutions

\BE  \htop_{r,s}^{BFK} = \pm {\ } \half \L(
      \L({\ctop-3\over3}\R) r + s \R)  \label{hrsB} \EE

\noi
and

\BE  \htop_k^{BFK} = {\ } \half 
      \L({\ctop-3\over3}\R) k  \label{hkB} \EE

\noi
where the superscript indicates that  these are U(1) charges
in the notation of BFK. Therefore we have to add $(\pm \half)$
to these expressions to translate them into our notation. Namely,

\BE  \htop_{r,s}^{BFK} + \half=
 \ccases{\half \L( \L({\ctop-3\over3}\R) r + s + 1 \R)}
{-\half \L( \L({\ctop-3\over3}\R) r + s - 1 \R)}   \EE

\BE  \htop_{r,s}^{BFK} - \half=
 \ccases{\half \L( \L({\ctop-3\over3}\R) r + s - 1 \R)}
{-\half \L( \L({\ctop-3\over3}\R) r + s + 1 \R)}   \EE

\noi
and

\BE  \htop_k^{BFK} + \half {\ } = {\ }
 \half \L( \L({\ctop-3\over3}\R) k + 1 \R) 
\EE

\BE  \htop_k^{BFK} - \half {\ } = {\ }
 \half \L( \L({\ctop-3\over3}\R) k - 1 \R) 
\EE

Comparing these expressions with the spectra given by \req{h0+rs},
\req{h0-rs}, \req{h1+rs} and \req{h1-rs} we see that the situation
is the same than in the Aperiodic algebra for Verma modules built
on chiral or antichiral primaries. Namely, half of the zeroes of 
$f_{r,s}^P=0$ correspond to uncharged singular states and the other
half correspond to charged singular states. However, since now
charged and uncharged singular states share the same spectra, the
zeroes of $g_k^P=0$ can be adjudicated to either of them
equivalently.

 Nevertheless, if we make the choice that adding $\half$
(or $-\half$) to the BFK spectra \req{hrsB} and \req{hkB} one gets
the spectra corresponding to the chirality $+$ (or $-$) singular 
states, one finds the following identifications. The upper solution
of $\htop_{r,s}^{BFK} + \half$ corresponds to the charged
singular states $\ket{\chi_R}_l^{(1)+}$  at level ${\ }r(s+2) \over2{\ }$,
 whereas the lower solution corresponds to the uncharged
singular states $\ket{\chi_R}_l^{(0)+}$ at level $rs\over2{\ }$.
The upper solution of $\htop_{r,s}^{BFK} - \half$  corresponds to
the uncharged singular states
 $\ket{\chi_R}_l^{(0)-}$ at level $rs\over2{\ }$,
while the lower solution corresponds to the charged
singular states $\ket{\chi_R}_l^{(-1)-}$
 at level $r(s+2)\over2{\ }$. The solutions $\htop_k^{BFK}+\half$
 and $\htop_k^{BFK}-\half$ correspond to the charged
singular states $\ket{\chi_R}_l^{(1)+}$ at level $k$, and 
$\ket{\chi_R}_l^{(-1)-}$ at level $(-k)$, respectively.
 Therefore, as happens with the Aperiodic NS 
algebra, the zeroes of the vanishing plane $g_k^P=0$ give the
 solutions  $\htop_{r,s}^{(1)+}$ and $\htop_{r,s}^{(-1)-}$
for $s=2$, while half of the zeroes of $f_{r,s}^P=0$ 
give the solutions for $s>2$.

\section{Conclusions and Final Remarks}\lvm

We have shown that
the determinant formulae for the N=2 Superconformal
algebra can be applied directly to incomplete Verma
modules, built on chiral primary states or on Ramond
ground states, provided one modifies appropiately the
interpretation of the zeroes in terms of identification
of the levels and relative charges of the singular states. 

We have written down the spectra of U(1) charges and conformal
 weights for the singular states of the Aperiodic NS algebra
  built on chiral and antichiral primary states and for the
singular states of the Periodic R algebra built on the 
Ramond ground states,
showing that, for those cases, the spectra 
corresponding to the charged singular states are
given by two-parameter expressions, like the spectra
corresponding to the uncharged singular states.
This is due to the fact that only half of the zeroes
of the quadratic vanishing surfaces
$f_{r,s}^A=0$ and $f_{r,s}^P=0$ correspond to
uncharged singular states, while the other half
of the zeroes correspond
to charged singular states, in contrast with
 the case of complete Verma modules for which all the
zeroes of $f_{r,s}^A=0$ and $f_{r,s}^P=0$ correspond
to uncharged singular states.

We have obtained these results simply by imposing the symmetries,
dictated by the spectral flows,
 between charged and uncharged singular states,
 starting with the ansatz that the zeroes of the ``chiral"
determinant formulae coincide with the zeroes of the determinant
formula of the Aperiodic NS algebra
 specialized to the cases $\Delta = \pm {\htop\over2}$.

Our results agree with all the known data (eight levels: from
level $\half$ until level 4). This, together with the simplicity
 of our derivation, strongly suggests that these results must be valid
at any level.

It is already remarkable the fact that the zeroes of the determinant 
formulae specialized to the cases $\Delta=\pm{\htop\over2}$ (or 
$\Delta={\ctop\over24}$), coincide with the zeroes of the
specific determinant formulae for the chiral 
representations of the Aperiodic algebra (or the representations of the 
Periodic algebra based on the Ramond ground states). The reason is that the 
highest weight equations resulting in the singular states,
and therefore in the zeroes of the determinant formulae, are different when 
one does or does not impose chirality on the primary
 states on which the singular 
states are built (or one does or does not impose 
that the Ramond primaries are annihilated 
by both $G_0^+$ and $G_0^-$). It seems that there
 exists a ``null vector conservation law"
or ``null vector transmutation" when switching
on and off chirality on the primary states, with charged
and uncharged singular states replacing each other.

We have also analyzed in very much detail the relation between the singular 
states of the Twisted Topological algebra and the singular states of the 
Aperiodic NS algebra, showing the direct relation between their 
corresponding spectra.
 As a result we have also derived the
spectrum of U(1) charges for
  the singular states of the Twisted Topological Algebra, which
agrees with the known data (until level 4). 

In addition, we have shown that the charged NS singular states
 built on chiral primaries have always relative charge $(-1)$,
while those built on antichiral primaries have always relative
charge $(+1)$. These states are mirrored to each other under the 
interchange $H_m \leftrightarrow -H_m,{\ \ }G_r^+ \leftrightarrow G_r^-$
because they are the two possible untwistings of the same topological 
singular state of type $\kc^{(1)G}$. In the same way, the charged
R singular states built on the Ramond ground states come only
in two types: $\ket{\chi_R}^{(1)+}$ and $\ket{\chi_R}^{(-1)-}$,
and they are also mirrored under 
$H_m \leftrightarrow -H_m,{\ \ }G_r^+ \leftrightarrow G_r^-$.

In the Appendix we have analyzed thoroughly the NS singular states 
$\ket{\chi_{NS}}_1^{(0)}$, $\ket{\chi_{NS}}_{3\over2}^{(1)}$ and 
$\ket{\chi_{NS}}_{3\over2}^{(-1)}$. We have written down the h.w. equations, 
with their solutions, for the cases of the primary states being non-chiral, 
chiral and antichiral, showing the ``null vector conservation law"
(or null vector transmutation) when 
switching on and off chirality on the primary states.

Finally, we have uncovered the existence of 
 non-highest weight singular states, which
are not secondary of any h.w. singular state. 

\vskip .2in

\centerline{\bf Acknowledgements}

We thank the Tata Institute of Fundamental Research (Bombay)
for hospitality,
where an important part of this paper was worked out. In
particular we thank S. Mukhi and A.K. Raina for very useful
 comments to this work.
We also thank E.B. Kiritsis and especially A.N. Schellekens
for many illuminating discussions and  Ch. Schweigert and
J. Gaite for help with the computer and for useful conversations.
Finally, we especially thank A. Kent for explaining to us that
our results were not in contradiction with the standard 
interpretation of the determinant formulae, as we had believed
in the previous version of this paper, and for many other valuable
comments to this work.

\setcounter{equation}{0}
\def\theequation{A.\arabic{equation}}

\subsection*{Appendix}

The general form of the charge $(+1)$ NS singular state at level 
${3\over2}$ is

\BE 
\ket{\chi_{NS}}_{3\over 2}^{(1)}=(\alpha L_{-1}G_{-1/2}^++
\beta H_{-1}G_{-1/2}^++\gamma G_{-3/2}^+)\ket{\Delta, \htop}\,.
\EE

\noi
The highest weight (h.w.) conditions
 $L_{m>0}\kc=H_{m>0}\kc=G_{r\geq\half}^+ \kc=
G_{r\geq\half}^- \kc=0$, which
 determine the coefficients ${\ }\alpha,{\ } \beta,{\ } \gamma, {\ }$ 
as well as the conformal weight $\Delta$  and the U(1) charge $\htop$
of the primary state, result as follows.
For $\ket{\Delta, \htop}$ non-chiral one obtains the 
equations

\begin{eqnarray}
\alpha(1+2\Delta)+\beta(1+\htop)+2\gamma &=& 0\nonumber\\
\alpha(1+\htop)+\beta{\ctop\over 3}+\gamma &=& 0\nonumber\\
(2\alpha+\beta)(2\Delta-\htop)+
\gamma(2\Delta-3\htop+2{\ctop\over3}) &=& 0\nonumber\\
\beta(2\Delta-\htop)-2\gamma &=& 0\nonumber\\
\alpha(2\Delta-\htop)+2\gamma &=& 0\nonumber\\
\alpha+\beta &=& 0\,.\label{eq3n}
\end{eqnarray}

\noi
We see that, for $\Delta\neq{\htop\over2}$, ${\ }\gamma$
must necessarily be different from zero 
(if $\gamma=0$ the whole vector vanishes). Hence
we can choose $\gamma=1$.
Solving for the other coefficients one obtains the solution, for 
$\Delta\neq{\htop\over2}$

\BE
\alpha={-2\over2\Delta-\htop}{\ \ }, \qquad \beta={2\over2\Delta-\htop}
\label{sol3n}\EE

\noi
with $\Delta-{3\over2}\htop+{\ctop-3\over3}=0$. This solution is given by 
the vanishing plane $g_{3/2}=0$, as one can check in eq. \req{gk}.

For the case $\Delta={\htop\over2}$ the solution is $\gamma=0$, $~\beta=-\alpha$ 
and $\htop={\ctop-3\over3}$. It is also given by the vanishing plane 
$g_{3/2}=0$. If we now specialize the general solution
\req{sol3n} to the case 
$\Delta=-{\htop\over2}$ we find

\begin{eqnarray}
 \alpha={6\over\ctop-3}{\ },\qquad\beta={6\over3-\ctop}{\ },\qquad
\htop={\ctop-3\over6}\,{\ }. \label{sol3na}
\end{eqnarray}

For $\ket{\Delta, \htop}$ chiral
(\ie\ $G_{-\half}^+ \ket{\Delta, \htop} =0,{\ } 
\Delta={\htop\over 2}$) the h.w. conditions only give the equation
 $\gamma=0$. Therefore $\ket{\chi_{NS}}_{3\over 2}^{(1)ch}$ vanishes
while $\ket{\chi_{NS}}_{3\over 2}^{(1)}$, for $\Delta={\htop\over2}$,
is a singular state with $\g=0,{\ }\b=-\a$, as we have just
 shown. This is a particular case
 of the general result that the charged singular states
 built on chiral primaries have always relative charge $q=-1$,
whereas those built on antichiral primaries have always $q=1$.  

For $\ket{\Delta, \htop}$
 antichiral (i.e. $G_{-\half}^- \ket{\Delta, \htop}=0,
{\ }\Delta=-{\htop\over 2}$) one gets the equations

\begin{eqnarray}
     \alpha(1-\htop)+\beta(1+\htop)+2\gamma &=& 0\nonumber\\
     \alpha(1+\htop)+\beta{\ctop\over 3}+\gamma &=& 0\nonumber\\
 (2\alpha+\beta)\htop+\gamma(2\htop-{\ctop\over 3}) &=& 0\nonumber\\
  \b\htop+\gamma &=& 0\nonumber\\
\alpha(1-\htop)+\beta+\gamma &=& 0\,.\label{eq3a}
\end{eqnarray}

\noi
Comparing these with equations \req{eq3n}, setting
 $\Delta=-{\htop\over2}$,
we see that here there is one equation less and the first four
equations coincide. The last equation here and the two last 
equations in \req{eq3n} are different. These equations
 correspond to the h.w. condition 
 $G_\half^-\ket{\chi_{NS}}_{3\over2}^{(1)}=0$.
As before $\gamma\neq 0$ necessarily, thus we set $\gamma=1$.
Solving for the other coefficients and for $\htop$ one 
obtains two solutions:

\BE
\a=\ccases{6\over\ctop-3}{\ctop-3\over6}\!,
\qquad\b=\ccases{6\over3-\ctop}{-1}\!,
\qquad\htop=\ccases{\ctop-3\over6}{1}\!. \label{sol3a}
\EE

\noi
The solution $\htop={\ctop-3\over 6}$ is the solution given by 
the vanishing plane $g_{3\over2}=0$, and therefore the only 
solution  for $\Delta =-{\htop\over2},{\ }\ket{\Delta,\htop}$
non-chiral, as we have shown in eq. \req{sol3na}. The solution
 $\htop=1$ corresponds to ${\ }\hat\htop_{1,2}{\ }$ in eq.\req{hh0rs},
 given by the vanishing quadratic surface $f_{1,2}=0$. These
two solutions are given, on the other hand, by
$~\htop^{(1)}_{2,2}~$ and $~\htop^{(1)}_{1,4}~$ in eq. \req{h1rs}.

\vskip .2in
\noi
Similarly, the general form of the charge $(-1)$ NS singular state at 
level ${3\over2}$ is

\BE
\ket{\chi_{NS}}_{3\over2}^{(-1)}=(\alpha L_{-1}G_{-1/2}^-+
\beta H_{-1}G_{-1/2}^-+\gamma G_{-3/2}^-) \ket{\Delta, \htop}\,.
\EE

\noi
Since this case is very similar to the previous one we will consider
only the chiral representations.
The h.w. conditions result in $\gamma=0$ for $\ket{\Delta, \htop}$ 
antichiral (therefore $\ket{\chi_{NS}}_{3\over2}^{(-1)a}$ vanishes),
 while for $\ket{\Delta, \htop}$ chiral one gets the equations

\begin{eqnarray}
    \alpha(1+\htop)+\beta(\htop-1)+2\gamma &=& 0\nonumber\\
    \alpha(\htop-1)+\beta{\ctop\over 3}-\gamma &=& 0\nonumber\\
(2\alpha-\beta)\htop+\gamma(2\htop+{\ctop\over 3}) &=& 0\nonumber\\   
\beta\htop+\gamma &=& 0\nonumber\\
 \alpha(\htop+1)-\beta+\gamma &=& 0\,,
\end{eqnarray}

\noi
where, as before, we can set $\gamma=1$. The other coefficients and the U(1) 
charge $\htop$ read

\BE
\a=\ccases{6\over\ctop-3}{\ctop-3\over6}\!,
\qquad\b=\ccases{6\over\ctop-3}{1}\!,
\qquad\htop=\ccases{3-\ctop\over6}{-1}\!.\label{sol3ch} \EE

\noi
These solutions correspond to
$(-\htop^{(1)}_{2,2})$ and $(-\htop^{(1)}_{1,4})$ respectively
(the upper solution is the solution given by the vanishing plane 
$g_{-{3\over2}}$). We see that $\ket{\chi_{NS}}_{3\over2}^{(1)a}$ and 
$\ket{\chi_{NS}}_{3\over2}^{(-1)ch}$ are mirrored under the interchange 
$H_m\leftrightarrow -H_m,{\ \ }G_r^+\leftrightarrow G_r^-$, reflecting the 
fact that they are the two different untwistings of the same topological 
singular state $\ket\chi_2^{(1)G}$, given by

\BE
\ket\chi_2^{(1)G}=(\cG_{-2}+\a\cL_{-1}\cG_{-1}+\b \cH_{-1}\cG_{-1})
\ket\phi_{\htop}
\EE

\noi
with

\BE
\a=\ccases{6\over\ctop-3}{\ctop-3\over6}\!,
\qquad\b=\ccases{6\over3-\ctop}{-1}\!,
\qquad\htop=\ccases{\ctop-3\over6}{1}\!. \EE

\noi
as the reader can verify using the twists \req{twa} and \req{twb}.

\vskip .15in

The general form of the uncharged NS singular state at 
level 1 is given by 

\BE
\ket{\chi_{NS}}_1^{(0)}=(\alpha L_{-1}+\beta H_{-1}+
\gamma G_{-1/2}^+G_{-1/2}^-)
\ket{\Delta, \htop}\,.
\EE

\noi
The h.w. conditions result as follows. For  
$\ket{\Delta, \htop}$ non-chiral one obtains the equations

\begin{eqnarray}
\a{\ } 2\Delta+\b{\ }\htop+\gamma(2\Delta+\htop) &=& 0\nonumber\\
\a{\ } \htop+\b{\ } {\ctop\over3}+\gamma(2\Delta+\htop) &=& 0\nonumber\\
\alpha-\beta-\gamma(2\Delta+\htop) &=& 0\nonumber\\
\alpha+\beta+\gamma(2\Delta-\htop+2) &=& 0\,.\label{1nch}
\end{eqnarray}

\noi
As before we can set $\gamma=1$ and we get

\BE
\alpha=\htop-1, \qquad \beta=-(2\Delta+1)\,,
\EE

\noi
with $\htop^2-2\Delta{\ctop-3\over3}-{\ctop\over3}=0$. This
solution corresponds to 
the quadratic vanishing surface $f_{12}=0$ in \req{frs}. It was
given before in \cite{Kir2} and \cite{YuZh}.

For $\Delta={\htop\over 2}$ the solutions are

\begin{eqnarray}
\alpha=\ccases{\ctop-3\over3}{-2}\!,{\ \ }
\hat\alpha=\ccases{\ctop+3\over3}{0}\!,{\ \ }
\beta=\ccases{-{\ctop+3\over3}}{0}\!,{\ \ }\htop=\ccases{\ctop\over3}{-1}\!,
\label{eq1nch} \end{eqnarray}

\noi
 $\a$ transforming into $\hat\a$ if we commute the term 
$G_{-\half}^+G_{-\half}^-\rightarrow G_{-\half}^-G_{-\half}^+$. These 
solutions correspond to $(-\htop_{1,2})$ and $(-\hat\htop_{1,2})$ in
\req{h0rs} and \req{hh0rs} respectively.

For $\Delta=-{\htop\over 2}$ the solutions are 

\begin{eqnarray}
\alpha=\ccases{-{\ctop+3\over3}}{0}\!,{\ \ }
\beta=\ccases{-{\ctop+3\over3}}{0}\!,{\ \ }\htop=\ccases{-{\ctop\over3}}{1}\!.
\label{eq1na} \end{eqnarray}

\noi
These solutions correspond to $\htop_{1,2}$ and $\hat\htop_{1,2}$ in 
eqns. \req{h0rs} and \req{hh0rs} respectively.

For the case of $\ket{\Delta, \htop}$
being chiral we can set $\gamma=0$ and the h.w. 
conditions  on $\ket{\chi_{NS}}_1^{(0)ch}$ give the equations
               
\begin{eqnarray}
\alpha+\beta &=& 0\nonumber\\
\a{\ }\htop+\b{\ }{\ctop\over3} &=& 0\,.
\end{eqnarray}

\noi
Comparing these with eqns. \req{1nch} 
for $\Delta={\htop\over2}$, we see that the first and fourth
equations in \req{1nch} coincide now with the first one here, 
 while the third equation in \req{1nch}, which corresponds
 to the h.w. condition $~G_\half^+\ket\chi=0$, disappears,
the reason being that the complete equation reads
$~(\alpha-\beta-\gamma(2\Delta+\htop))G_{-\half}^+\ket{\Delta,\htop}=0$.
The solution to these equations is
${\ }\beta=-\alpha,{\ \ \ }\htop={\ctop\over 3}{\ }$, that is

\BE \ket{\chi_{NS}}_1^{(0)ch}= (L_{-1} - H_{-1})
{\ \ }\ket{\Delta={\ctop\over6}, {\ } \htop={\ctop\over3}}\,. \EE

\noi
Therefore, only the solution $(-\htop_{1,2})={\ctop\over3}$
 remains after switching
on chirality on the primary state $\ket{\Delta,\htop}$, while
$(-\hat\htop_{1,2})=-1$ is not a solution anymore. Not only that, but
in addition the singular state $\ket{\chi_{NS}}_1^{(0)}$
vanishes for $(-\hat\htop_{1,2})$ when one imposes
chirality on $\ket{\Delta,\htop}$, as one can check in \req{eq1nch}.

For the case of $\ket{\Delta, \htop}$
being antichiral the h.w. 
conditions on $\ket{\chi_{NS}}_1^{(0)a}$ give the equations
               
\begin{eqnarray}
\alpha-\beta &=& 0\nonumber\\
\a{\ }\htop+\b{\ }{\ctop\over3} &=& 0\,.
\end{eqnarray}

\noi
Comparing these with eqns. \req{1nch} we see that now the 
last equation in \req{1nch}, which corresponds
 to the h.w. condition $~G_\half^-\ket\chi=0$, has disappeared,
the reason being that the complete equation reads
$~(\alpha+\beta+\gamma(2\Delta-\htop+2))
G_{-\half}^-\ket{\Delta,\htop}=0$.
The solution to these equations is
${\ }\beta=\alpha,{\ \ \ }\htop=-{\ctop\over 3}{\ }$, that is

\BE \ket{\chi_{NS}}_1^{(0)a}= (L_{-1} + H_{-1})
{\ \ }\ket{\Delta={\ctop\over6}, {\ } \htop=-{\ctop\over3}}\,. \EE

\noi
Therefore, only the solution ${\ }\htop_{1,2}=-{\ctop\over3}{\ }$ 
remains after switching
on antichirality on the primary state $\ket{\Delta,\htop}$, while
the solution ${\ }\hat\htop_{1,2}=1{\ }$ disappears. In addition, as
happened in the chiral case,
 the singular state $\ket{\chi_{NS}}_1^{(0)}$
vanishes for ${\ }\hat\htop_{1,2}{\ }$ when one imposes
antichirality on $\ket{\Delta,\htop}$,
 as one can check in \req{eq1na}.

Observe that $\ket{\chi_{NS}}_1^{(0)ch}$ and
$\ket{\chi_{NS}}_1^{(0)a}$ are symmetric under the interchange
$H_m \leftrightarrow -H_m, {\ } G^+_r \leftrightarrow G^-_r{\ }$,
as expected. This also happens for the singular states
$\ket{\chi_{NS}}_1^{(0)}$ specialized to the cases
 $\Delta = {\htop\over2}$ and $\Delta = -{\htop\over2}$,
eqns. \req{eq1na} and \req{eq1nch}, as one can check (one 
has to take into account that the
commutation $G^+_{-1/2} G^-_{-1/2} \rightarrow G^-_{-1/2} G^+_{-1/2}$
produces a global minus sign).

We have seen that the uncharged states $\ket{\chi_{NS}}_1^{(0)}$,
for $\htop=\hat\htop_{1,2} =1{\ \ } 
(\htop=- \hat\htop_{1,2} =-1{\ }), {\ }\Delta = -\half$,
vanish when one switches on antichirality (chirality) respectively
on 
$\ket{\Delta, \htop}$. An interesting related question now is what
happens with the charged singular states
 $\ket{\chi_{NS}}_{3\over2}^{(1)a}$ for $\htop= \hat\htop_{1,2}=1{\ }$
and $\ket{\chi_{NS}}_{3\over2}^{(-1)ch}$ for $\htop=-\hat\htop_{1,2}=-1{\ }$
when one switches off antichirality or chirality on
$\ket{\Delta, \htop}$; that is, which kind of states these singular
states become once they are not h.w. singular states anymore.
One might think that these h.w. charged singular states which disappear at
level ${3\over2}$ turn into secondary singular states of the h.w. uncharged
singular states which appear at level 1. However this is not quite
true because, as we will see, these charged states, which are still
singular although not h.w., can descend down to the h.w. uncharged
singular states, but not the other way around.

To see this 
in some detail
let us write the h.w. uncharged singular states at level 1
for $\htop=\pm 1 {\ }, \Delta= \mp {\htop\over2}=-{1\over2}{\ },$
given by eqns. \req{eq1na} and \req{eq1nch}. 
 For $\htop=1,{\ }\Delta=-{1\over2}{\ },$ the singular state is 

\BE \ket{\chi_{NS}}_1^{(0)} = G^+_{-1/2} G^-_{-1/2}{\ \ }
    \ket{\Delta=-{\half},{\ }\htop=1}\,. \label{v1n1} \EE

\noi
Its level ${3\over2}$ descendant with relative charge $q=1$ vanishes
since ${\ }G^+_{-1/2}{\ }\ket{\chi_{NS}}_1^{(0)} =0{\ }$.
Similarly, for
$\htop=-1,{\ }\Delta=-{1\over2}{\ },$ the singular state is 

\BE \ket{\chi_{NS}}_1^{(0)} = G^-_{-1/2} G^+_{-1/2}{\ \ }
    \ket{\Delta=-{\half},{\ }\htop=-1}\,, \label{v1n-1} \EE

\noi
and its level ${3\over2}$ descendant with relative charge $q=-1$ vanishes.
(It is also straightforward to see that the uncharged singular
states \req{v1n1} and \req{v1n-1} vanish when one imposes
antichirality and chirality, respectively, on the primary states,
as we said before.)

On the other hand, the h.w. charged singular states for 
$\htop=\pm 1 {\ }, \Delta= \mp {\htop\over2}=-{1\over2}{\ },$  
given by \req{sol3a} and \req{sol3ch}, are

\BE 
\ket{\chi_{NS}}_{3\over 2}^{(1)a}=
({\ctop-3\over6}{\ } L_{-1}G_{-1/2}^+ -
 H_{-1}G_{-1/2}^+{\ }+{\ } G_{-3/2}^+)
{\ \ }\ket{\Delta=-\half,{\ } \htop=1}  \label{v3a1}
\EE
 
\noi
and

\BE 
\ket{\chi_{NS}}_{3\over 2}^{(-1)ch}=
({\ctop-3\over6}{\ } L_{-1}G_{-1/2}^- +
 H_{-1}G_{-1/2}^-{\ }+{\ } G_{-3/2}^-)
{\ \ }\ket{\Delta=-\half,{\ } \htop=-1}  \label{v3c-1}
\EE

If we now switch off antichirality (chirality) on the primary
state $\ket{\Delta,\htop}$, then the h.w. condition
$G^-_{1/2} \ket{\chi_{NS}}_{3\over 2}^{(1)a}=0{\ }$
($G^+_{1/2} \ket{\chi_{NS}}_{3\over 2}^{(-1)ch}=0{\ }$)
is not satisfied anymore, although the states are still singular,
\ie\ have zero norm, as the reader can verify. However, these 
states cannot be secondary states of the uncharged 
singular states \req{v1n1} and \req{v1n-1}, as we have just discussed,
nor are they secondary states of any level $\half$ singular states.
In other words, $\ket{\chi_{NS}}_{3\over 2}^{(1)a}$ and
$\ket{\chi_{NS}}_{3\over 2}^{(-1)ch}$ become non-highest weight
singular states, not secondary of any singular states, once we
switch off antichirality and chirality on the primary state
$\ket{\Delta,\htop}$, respectively. 

Nevertheless, these level $3\over2$ zero-norm states do descend to the
level 1 uncharged singular states \req{v1n1} and \req{v1n-1}
under the action of $G^-_{1/2}{\ }$ and $G^+_{1/2}{\ }$,
 respectively, as is easy to check, but not the other
way around. The reason is that the singular
states \req{v1n1} and \req{v1n-1} do not build complete Verma
modules, as is the usual case for the h.w. singular states of
the N=2 superconformal algebra. 
 
\vskip .15in
To finish let us discuss further the mechanism of
 ``null vector conservation" or ``transmutation". For this it
is instructive to analyze the behaviour of the uncharged singular state
$\ket{\chi_{NS}}_1^{(0)}$  and its level $3\over2$ 
descendants near the limits $\htop \rightarrow \pm1, {\ }
\Delta \rightarrow - {\half}$.

Let us start with  $\htop$ near 1. Thus we set
${\ }\htop=1+\epsilon{\ },{\ }\Delta=-{1\over 2}(1+\delta){\ },$
$\Delta$ and $\htop$ satisfying the quadratic vanishing
surface relation, which results in

\BE {\ctop\over3}={\delta-2\epsilon-\epsilon^2\over\delta}\,. \label{ced} \EE

\noi
The state $\ket{\chi_{NS}}^{(0)}_1$   is expressed now as

 \BE
\ket{\chi_{NS}}_1^{(0)}=(\epsilon L_{-1}+\delta H_{-1}+
G_{-1/2}^+G_{-1/2}^-)\ket{\Delta,\htop}\,.
\EE

\noi
Its charge $(+1)$ descendant at level $3\over2$, which we denote as
$\ket\Ups^{(1)}_{3\over2}$, is a singular state which, in 
principle, is not h.w.

$$\ket\Ups^{(1)}_{3\over2} = 
G_{-1/2}^+ \ket{\chi_{NS}}_1^{(0)} =
(\epsilon L_{-1}G_{-1/2}^++\delta H_{-1}G_{-1/2}^+
-\delta G_{-3/2}^+)\ket{\Delta,\htop}\,. $$

\noi
Now comes a subtle point. In principle
we can normalize $\ket\Ups^{(1)}_{3\over2}$ in the same way as
$\ket{\chi_{NS}}^{(1)a}_{3\over2}$, \ie\ dividing all the
coefficients by $(-\delta)$ so that the coefficient
of $G^+_{-3/2}$ is 1, resulting in

\BE \ket\Ups^{(1)}_{3\over2} = 
((-{\epsilon\over\delta}) L_{-1}G_{-1/2}^+-H_{-1}G_{-1/2}^+
+G_{-3/2}^+)\ket{\Delta,\htop}\,.  \label{chied}
\EE
 
\noi
However, these two normalizations are not
equivalent when taking the limit $(\htop=1, \Delta=-\half){\ }$,\ie\ 
${\ }(\e \rightarrow 0, \delta \rightarrow 0){\ }$. Namely
 $\ket\Ups^{(1)}_{3\over2}$ vanishes with the first
normalization whereas with the second normalization
it becomes the h.w. singular state  $\ket{\chi_{NS}}^{(1)a}_{3\over2}{\ }$, 
since $(-{\epsilon\over\delta})$ turns into
$({\ctop-3\over6})$, as can be deduced easily from \req{ced}.                       

It seems that the two normalizations distinguish whether the
primary state $\ket{\Delta,\htop}$ approaches an antichiral or a
 non-chiral state as $\htop \rightarrow 1, \Delta 
\rightarrow -\half$. This is indeed true, the reason is that if we normalize
$\ket{\chi_{NS}}^{(0)}_1$ in the same way as its descendant
$\ket\Ups^{(1)}_{3\over2}$, then in the
second normalization, \ie\ dividing by $(-\delta)$, it blows up
approaching the limit $(\e \rightarrow 0, \delta \rightarrow 0)$
unless the term $G^+_{-1/2} G^-_{-1/2}$ goes away, exactly what
happens if $\ket{\Delta,\htop}$ is antichiral. 
But the resulting uncharged state without the term
$G^+_{-1/2} G^-_{-1/2}$  is not a
 singular state anymore. The action
of $G^+_{-1/2}$ on this state results
 precisely in the charged h.w. singular state
 $\ket{\chi_{NS}}^{(1)a}_{3\over2}{\ }$. 

We see therefore that, in the limit  $(\htop=1, \Delta=-\half){\ }$,
 the first normalization produces the h.w. uncharged singular state
at level 1 
${\ }\ket{\chi_{NS}}^{(0)}_1$, with a vanishing charge (+1) descendant
at level $3\over2$, whereas the second normalization produces the h.w.
charge (+1) singular state
at level $3\over2$ ${\ }\ket{\chi_{NS}}^{(1)a}_{3\over2}{\ }$.

Repeating this analysis for the case $\htop$ near $-1$
we find that
in the limit  $(\htop=-1, \Delta=-\half){\ }$,  
 the first normalization produces the h.w. uncharged singular state
at level 1 
${\ }\ket{\chi_{NS}}^{(0)}_1$, with a vanishing charge (-1) descendant
at level $3\over2$, whereas the second normalization produces the h.w.
charge (-1) singular state
at level $3\over2$ ${\ }\ket{\chi_{NS}}^{(-1)ch}_{3\over2}{\ }$.

\end{document}